\documentclass[preprint,onecolumn,amsmath,amssymb,floatfix]{revtex4-1}
\usepackage{graphicx}
\usepackage{times}
\usepackage{color}

\begin{document}

\title{Optical Rectification and Field Enhancement in a Plasmonic Nanogap}

\author{Daniel R. Ward,$^{1}$ Falco H{\"u}ser,$^{2}$ Fabian Pauly,$^{2}$
Juan Carlos Cuevas,$^{3}$ Douglas Natelson$^{1,~4,~\ast}$}

\affiliation{$^{1}$ Department of Physics and Astronomy, Rice University, 6100 Main St., Houston, TX, USA}
\affiliation{$^{2}$ Institut f{\"u}r Theoretische Festk{\"o}rperphysik, Karlsruhe Institute of Technology, 76131 Karlsruhe, Germany}
\affiliation{$^{3}$ Departamento de F\'{\i}sica Te\'orica de la Materia Condensada, Universidad Aut\'onoma de Madrid, 28049 Madrid, Spain}
\affiliation{$^{4}$ Department of Electrical and Computer Engineering, Rice University, 6100 Main St., Houston, TX, USA}

\begin{abstract}
{\bf Metal nanostructures act as powerful optical antennas\cite{Muhlschlegel2005,Schuck2005} because collective modes of the electron fluid in the metal are excited when light strikes the surface of the nanostructure.  These excitations, known as plasmons, can have evanescent electromagnetic fields that are orders of magnitude larger than the incident electromagnetic field.  The largest field enhancements often occur in nanogaps between plasmonically active nanostructures\cite{Jiang2003,Li2003}, but it is extremely challenging to measure the fields in such gaps directly.  These enhanced fields have applications in surface-enhanced spectroscopies\cite{Otto1992,Hartstein1980,Fort2008}, nonlinear optics\cite{Muhlschlegel2005,danckwertsPRL2007,bouhelier2003,Ghenuche2008}, and nanophotonics\cite{Akimov2007,Dionne2009,Yu2007,Oulton2009,Noginov2009}.  Here we show that nonlinear tunnelling conduction between gold electrodes separated by a subnanometre gap leads to optical rectification, producing a DC photocurrent when the gap is illuminated.  Comparing this photocurrent with low frequency conduction measurements, we determine the optical frequency voltage across the tunnelling region of the nanogap, and also the enhancement of the electric field in the tunnelling region, as a function of gap size.  The measured field enhancements exceed 1000, consistent with estimates from surface-enhanced Raman measurements\cite{wardNanoLett2007,wardNanoLett2008,WardJPCM2008}.  Our results highlight the need for more realistic theoretical approaches that are able to model the electromagnetic response of metal nanostructures on scales ranging from the free space wavelength, $\lambda$, down to $\sim \lambda/1000$, and for experiments with new materials, different wavelengths, and different incident polarizations.} 
\end{abstract}

\noindent Published as D. R. Ward, F. H{\"u}ser, F. Pauly, J. C. Cuevas, and D. Natelson, {\it Nature Nano.} {\bf 5}, 732-736 (2010), http://dx.doi.org/10.1038/nnano.2010.176

\maketitle

The probability of an electron being able to tunnel across a small gap decays exponentially with distance, so electron tunnelling can be used to measure distances with high precision, as happens in scanning tunnelling microscopy (STM).  Moreover, nonlinearity in tunnelling conduction can convert the optical frequency response of the metal into a DC current.  This rectification process, originally proposed as a means of inferring the tunnelling time\cite{cutlerPRB1987}, allows the optical frequency voltage difference across the tunnelling gap to be determined.  Evidence for microwave rectification\cite{tuJCP2006} and optical rectification\cite{nguyenElectronDevices1989,bragasAPL1998} has been reported for STM experiments, but thermal and surface photovoltage effects\cite{tuJCP2006} complicate interpretation of the optical data.


Suppose that the DC two-terminal current-voltage characteristics of some object are $I(V_{\mathrm{DC}})$, where $I$ is the current and $V_{\mathrm{DC}}$ is the DC voltage bias.  If the underlying conduction mechanism remains valid at some frequency scale $2\omega$, then in the presence of some small AC voltage bias, $V_{\mathrm{AC}} \cos(\omega t)$, in addition to $V_{\mathrm{DC}}$, one expects\cite{tuJCP2006} a total current of 
\begin{eqnarray}
\label{eq:classicalrect}
I & \approx & I(V_{\mathrm{DC}}) + \frac{\partial I}{\partial V}|_{V_{\mathrm{DC}}} V_{\mathrm{AC}}\cos (\omega t) + \frac{1}{2}\frac{\partial^2 I}{\partial V^2}|_{V_{\mathrm{DC}}} V_{\mathrm{AC}}^2 \cos^2 (\omega t) + ... \\
& \approx & \left[I(V_{\mathrm{DC}}) + \frac{1}{4}\frac{\partial^2 I}{\partial V^2}|_{V_{\mathrm{DC}}} V_{\mathrm{AC}}^2\right]
 + \frac{\partial I}{\partial V}|_{V_{\mathrm{DC}}} V_{\mathrm{AC}}\cos(\omega t) - \frac{1}{4}\frac{\partial^2 I}{\partial V^2}|_{V_{\mathrm{DC}}} V_{\mathrm{AC}}^2 \cos  (2\omega t) + .... \nonumber
\end{eqnarray}

In addition to $I(V_{\mathrm{DC}})$, in the presence of the AC excitation there is a rectified DC current proportional to the nonlinearity of the conductance and the square of the AC excitation amplitude.  As demonstrated\cite{tuJCP2006} using STM and microwave irradiation, one may use a low frequency AC voltage and lock-in amplifier to measure $\partial^2 I/\partial V^2$ using the second harmonic of the AC voltage.  Using lock-in techniques to measure the DC current due to microwave irradiation of a known microwave voltage, Tu \textit{et al.}\cite{tuJCP2006} find quantitative agreement between the rectified microwave current and the expectations of Eq.~(\ref{eq:classicalrect}).  

In the optical case, \emph{in the limit that this rectification picture is valid}, one would expect the radiation-induced AC response of the metal to lead to a photocurrent, $I_{\mathrm{photo}}$, in addition to the usual DC current, $I(V_{\mathrm{DC}})$.  One may then infer the local, radiation-induced AC voltage, $V_{\mathrm{opt}}$,  by comparing $I_{\mathrm{photo}}$ and detailed low frequency characterization of the conductance.  In a tunnel junction, the differential conductance, $\partial I/\partial V$, at zero bias can be related to the interelectrode separation through the exponential decay of the conductance with increasing distance, $G \sim \exp(-\beta(d-d_{0}))$.  Here $d$ is the internuclear distance between closest atoms, $d_{0}$ is the size of the metal lattice constant, and $\beta$ is the attenuation factor.  The \emph{average local electric field at the closest interelectrode separation}, $V_{\mathrm{opt}}/(d-d_{0})$, may then be inferred and compared with the incident field known from the incident optical intensity, giving a quantitative measure of the field enhancement due to the metal optical response.   The $V_{\mathrm{opt}}$ inferred experimentally is that experienced by the optical-frequency tunnelling electrons.  Since both $V_{\mathrm{opt}}$ and $d-d_{0}$ are found from tunnelling, a very local process, this is self-consistent.   Voltage drops within the metal far from the tunnelling gap are not relevant.  However, there are subtleties to consider, including the role of screening in the metal and the locality of the tunnelling process.  We discuss this further in Supplemental Information.  

A quantum treatment of radiation interacting with a nanoscale junction involves photon-assisted tunnelling\cite{Viljas2007} and has been applied to atomic-scale metal contacts\cite{guhrPRL2007,ittahNL2009}. When a junction is illuminated with radiation of energy $\hbar \omega$, the junction's plasmonic response results in an AC voltage, $V_{\mathrm{opt}}$ at frequency $\omega$ across the junction.  Following the Tien-Gordon approach\cite{tienPR1963}, the quantum correction to the DC current induced by the radiation in the limit of small AC amplitudes ($eV_{\mathrm{opt}} \ll \hbar\omega$) is given by
\begin{equation}
I(V_{\mathrm{DC}},V_{\mathrm{opt}},\omega)-I(V_{\mathrm{DC}}) = \frac{1}{4}V_{\mathrm{opt}}^{2}
\left[ \frac{I(V_{\mathrm{DC}}+\hbar\omega/e) - 2 I(V_{\mathrm{DC}}) + I(V_{\mathrm{DC}}-\hbar \omega/e)}{(\hbar \omega/e)^2}\right].
\label{eq:tiengordon}
\end{equation}
If the tunnelling nonlinearity is small on the voltage scale $\hbar \omega/e$, this expression reduces to Eq.~(\ref{eq:classicalrect}) with $V_{\mathrm{opt}}$ playing the role of $V_{\mathrm{AC}}$.  This occurs when the transmittance of the junction $\tau(E)$ is smooth as a function of energy near the Fermi level of the leads, $E_{\mathrm{F}}$, throughout $E_{\mathrm{F}}\pm \hbar \omega$.


Details of the sample fabrication and the optical and electronic measurements are described in Methods.  Nanogaps are formed in the Au constrictions via electromigration\cite{ParkAPL1999}.  The electromigration process is stopped once the zero bias DC conductance of the gaps is less than $2e^2/h \equiv G_0$.  Typical conductances in this experiment are on the order of 0.6~$G_0$ to 0.016~$G_0$ indicating average gaps, $d-d_{0}$, ranging from 0.03~nm to 0.23~nm.  Figure 1 shows the measurement system and basic characterization of a representative nanogap.  The illumination wavelength is 785~nm, with a peak intensity of 22.6 kW/cm$^{2}$, and measurements are performed in vacuum at 80~K.

The DC $I-V$ characteristics of the samples after electromigration are
weakly nonlinear, with no sharp features, as expected for clean vacuum
tunnel junctions.  These junctions, ideally devoid of molecules,
show \textit{no} molecular surface-enhanced Raman scattering
signal\cite{wardNanoLett2007,wardNanoLett2008}, as expected.  Fig.~2 shows
representative $\partial^{2}I/\partial V^2$ vs. $V_{\mathrm{DC}}$
curves and simultaneously recorded $I_{\mathrm{photo}}$
vs. $V_{\mathrm{DC}}$ data for three different samples, acquired with
the beam fixed over the nanogap.  For each set of curves shown, the
amplitude of $V_{\mathrm{AC}}$ has been set such that
$\frac{1}{4}V_{\mathrm{AC}}^{2}\partial^{2}I/\partial V^2$ and
$I_{\mathrm{photo}}$ are the same amplitude.  The fact that
$I_{\mathrm{photo}}$ is directly proportional to
$\partial^{2}I/\partial V^2$ consistently over the whole DC bias
range, including through the $V_{\mathrm{DC}}$ where both change sign,
strongly implies that the photocurrent originates from the
rectification mechanism.  In this case, when
$\frac{1}{4}V_{\mathrm{AC}}^{2}\partial^{2}I/\partial V^{2} =
I_{\mathrm{photo}}$, this implies $V_{\mathrm{AC}} =
V_{\mathrm{opt}}$, where $V_{\mathrm{opt}}$ is the amplitude of the
optically induced voltage between the two electrodes that
oscillates at $\omega/2 \pi = 3.8 \times 10^{14}$~Hz.  When the
rectification mechanism is responsible for the photocurrent,
comparison with low frequency electrical measurements gives a means of
quantitatively assessing the plasmon-induced AC voltage
across the nanoscale interelectrode gap.  As a consistency check, we
have mapped the optically rectified current, $I_{\mathrm{photo}}$, at
fixed $V_{\mathrm{DC}}$ (while monitoring $I$, $\partial I/\partial
V$, and $\partial^{2}I/\partial V^{2}$, to ensure junction stability
during the mapping procedure).  As shown in Fig.~1e,
$I_{\mathrm{photo}}$ is only produced when the gap itself is
illuminated.

Previous tunnelling optical rectification experiments have been hampered by thermal voltages and thermal expansion.  We do not expect thermal voltages in our experiment since the device is a comparatively symmetric design, and with a centered laser spot there should be no temperature gradient between the electrodes, even in the presence of minor structural asymmetries.  The macroscale unilluminated parts of the electrodes also act as thermal sinks.  Differential thermal voltages caused by asymmetry of illumination would show up as systematic trends in $I_{\mathrm{photo}}$ as a function of laser position, and Fig.~1e shows no such trends.  Thermal expansion should be irrelevant, since the electrodes are intimately mechanically coupled to the substrate, with essentially no unsupported electrode material.  To check for thermal effects, we looked for changes in $\partial I/\partial V$ at $\omega_{chopper}$, since any laser-induced expansion should modulate the interelectrode gap and thus the conductance.  No measurable effect was found to better than the $10^{-4}$ level.

Optical rectification is further supported by 
measurements of $I_{\mathrm{photo}}$ vs. incident laser
power.  In Fig.~3 we see the expected linear dependence.  The best fit
line has a slope of 0.37~nA/(kW/cm$^2$).  Further, as expected,
$V_{\mathrm{DC}}$ can be selected such that $\partial^{2}I/\partial
V^2$ is zero (not always at $V_{\mathrm{DC}}=0$), and at that bias
$I_{\mathrm{photo}}$ measurements show no detectable current.

The validity of the rectification picture is not completely obvious
for this system \textit{a priori}.  Detailed calculations of the
transmission as a function of energy,
as shown in Fig.~4, demonstrate that for this particular 
material and illuminating wavelength, corrections to
the classical rectification picture should be relatively small.  At
higher DC biases ($>$ 100 mV, not shown), we do often observe
deviations of the photocurrent from the measured
$(1/4)V_{\mathrm{AC}}^{2}\partial^{2}I/\partial V^{2}$.  

To infer the nanogap size from the conductance measurements, we have
performed theoretical simulations of the breaking process of gold
atomic contacts.  In these simulations, we have used density
functional theory to determine the geometry of the junctions as well
as to compute the transport properties within the Landauer approach,
as detailed in Ref.~\cite{Pauly2008}.  In this approach the
low-temperature linear conductance is given by $G = G_0
\tau(E_{\mathrm{F}})$, where $\tau(E_{\mathrm{F}})$ is the
transmission of the contact evaluated at the Fermi energy,
$E_{\mathrm{F}}$.  To simulate the rupture of the gold nanowires, we
start out with a single-atom contact (\textit{e.g.,}
Fig.~4a).  We separate the electrodes stepwise, re-optimize the
geometry in every step, and compute the
corresponding conductance.  As in Fig.~4c, in the relevant range, the linear conductance decays exponentially with an attenuation factor
of $\beta \approx 1.8$ \AA$^{-1}$. Extensive simulations (Supplementary Information) show that
this exponent varies within 10\% depending on the geometry of
the contact.

Using $\beta = 1.85$ \AA$^{-1}$ (averaged over different junction 
structures), the interelectrode gap, 
$d-d_{0}$, can be found from the measured  
$\partial I/\partial V$.  With this and the inferred $V_{\mathrm{opt}}$, 
the enhanced electric field local
to the tunnelling gap may be determined.  The experimental uncertainty
in $V_{\mathrm{opt}}$ is approximately 5\%, based on the comparisons
shown in Fig.~2.  The experimental uncertainty in the tunnelling
conductance at zero bias is  $\sim$~1\%.  The dominant
uncertainty in the inferred interelectrode separation is systematic,
due to variation of $\beta$ with morphology of the
metal surfaces. Assuming that $d - d_{0}$ is the
appropriate scale for computing the local electric field, for
the samples in Fig.~2, the optical electric fields (RMS) are
$2.1 \times 10^8$~V/m, $5.7 \times 10^8$~V/m, and $3.6 \times
10^8$~V/m, respectively.  Based on gaussian beam shape and incident power,
the free-space incident RMS field is $2.9 \times 10^5$~V/m.  Therefore we
determine local enhancement factors of the electric field to be
718, 1940, and 1230.  These very large enhancements exceed
expectations from idealized FDTD calculations\cite{wardNanoLett2007}
for wider gaps, and are consistent with observations of
single-molecule Raman emission in such junctions
\cite{wardNanoLett2008,WardJPCM2008}.  At sufficiently high
electric field strengths the gap geometry can become unstable.  We
believe that this is the reason why $I-V$ curves of some devices
exhibit instabilities.  In the absence of the optical field 
or at considerably reduced incident laser fluences, the $I-V$ curves
are very stable.

Our calculations clarify why the classical rectification formula works
in our case.  The transmission function (Fig. 4d) of these gold contacts shows only a weak energy dependence around the Fermi energy in the interval $[E_{\mathrm{F}} - \hbar \omega,E_{\mathrm{F}} + \hbar \omega]$. This indicates that the quantum result of Eq. (\ref{eq:tiengordon}) reduces to the classical rectification formula of Eq. (\ref{eq:classicalrect}).  This smoothness is because
transport in gold contacts is dominated by the $s$ atomic
orbitals, with a relatively flat density of states.  Similar physics
should therefore be expected in other $s$-valent metals, such as Ag.

In several devices, conductance decreased over the course of hours of
measurements, as the metal electrodes rearranged themselves, allowing
us to track variations in $V_{\mathrm{opt}}$ (and inferred enhanced
electric field) as a function of interelectrode separation (Fig.~5).
Despite device-to-device variability, the general trend shows
increasing enhancement with decreasing $d-d_{0}$.  This trend is
slower than $1/(d-d_{0})$ and weaker than classical
expectations\cite{zuloagaNanoLett2009}.  Such a comparatively weak
increase in enhancement at low $d-d_{0}$ is qualitatively consistent
with calculations that examine quantum tunnelling corrections to
plasmonic enhancements in nanoparticle
dimers\cite{zuloagaNanoLett2009,maoAPL2009}.  However, those
calculations predict an actual \emph{decrease} in enhancement when
$d-d_{0}$ is small enough that interelectrode conductance is strong
enough to short out the nanogap plasmons.  From our data and previous
observations\cite{wardNanoLett2008}, this decrease in enhancement does
not happen in gold nanojunctions until interelectrode conductance
exceeds $G_{0}$.  Our measurements highlight the need for detailed,
realistic theoretical treatments of these nanostructures,
incorporating optical interactions, quantum effects, and dynamical
screening.

\noindent {\bf Methods}

Samples are fabricated using standard electron-beam lithography
techniques to define metal (1 nm Ti/15 nm Au) constrictions 100 nm in
width and 600 nm in length between larger contact pads, on highly
doped n-type Si substrates covered with 200 nm of thermally grown
SiO$_2$.  After cleaning in oxygen plasma, devices are wirebonded to a
chip carrier, installed in a microscope cryostat, and cooled to 80 K.

After electromigration the sample is illuminated with a 785~nm laser
focused to a gaussian spot with FWHM 1.9~$\mu$m and a peak intensity
of 22.6~kW/cm$^2$.  The laser is rastered over the sample surface and
the Raman response of Si substrate is measured at 520~cm$^{-1}$.  The
Au electrodes attenuate the Si Raman emission, allowing precise
mapping of the nanogap structure as shown in Fig. 1d.  Once the
nanogap is located via the Si Raman map, the laser is centered over
the gap.  The laser polarization is aligned along the original
constriction.

The sample is electrically characterized via the apparatus presented
in Fig.~1a using two lock-in amplifiers and a current amplifier.  The
first lock-in amplifier applies an AC excitation ($V_{\mathrm{AC}}
\cos \omega_{1} t$) to the sample at $\omega_{1}$ = 2.0~kHz and
simultaneously measures the first ($\propto \partial I/\partial V$)
and second harmonic response ($=\frac{1}{4} V_{\mathrm{AC}}^{2}
\partial^{2}I/\partial V^2$) of the nanogap to this excitation.  There
is no measurable change in either of these responses when the laser
illumination is added or removed.  The first harmonic response is
essentially in phase with the $\omega_{1}$ reference.  The second
harmonic response is phased relative to the $2\omega_{1}$ reference by
180 degrees as defined by the lock-in, such that the integrated
$\partial^{2}I/\partial V^2$ signal gives back the $ \partial
I/\partial V$ signal.  The second lock-in amplifier is referenced to
(and in phase with) an optical chopper at $\omega_{chopper}$ = 232~Hz
and measures the current generated due to the radiation
($I_{\mathrm{photo}}$).  A summing amplifier is used to add a DC
voltage $V_{\mathrm{DC}}$ to $V_{\mathrm{AC}}$.  With this
arrangement, $I(V_{\mathrm{DC}})$, $\partial I/\partial V$,
$\partial^{2}I/\partial V^2$, and $I_{\mathrm{photo}}$ as a function
of $V_{\mathrm{DC}}$ can be measured simultaneously.

\noindent {\bf [Supplementary Information]} is linked to the online version of the paper at www.nature.com/nature.
\\

\noindent {\bf[Acknowledgments]}  D.N. and D.R.W. acknowledge support by Robert A. Welch Foundation grant C-1636 and the  Lockheed Martin Advanced Nanotechnology Center of Excellence at Rice (LANCER).  F.H. and J.C.C. acknowledge support of the DFG, the Baden-W\"urttemberg Siftung, the EU through the BIMORE network (grant MRTN-CT-2006-035859), and the Spanish MICINN (grant FIS2008-04209).  FP acknowledges funding of a Young 
Investigator Group.
\\

\noindent{\bf [Author Contributions]}  D.R.W. fabricated the devices, performed all measurements, and analyzed the data.  D.N. supervised and provided continuous guidance for the experiments and the analysis.  F.P., F.H., and J.C.C. carried out the theoretical modeling and DFT calculations.  The bulk of the paper was written by D.R.W. and D.N.  All authors discussed the results and contributed to manuscript revision.
\\

\noindent{\bf [Competing Interests]}  The authors declare that they have no competing financial interests.
\\

\noindent{\bf[Correspondence]}  Correspondence and requests for materials should be addressed to D.N.~(email: natelson@rice.edu).

\clearpage

\begin{figure}
\begin{center}
\includegraphics[width=8.5cm,clip]{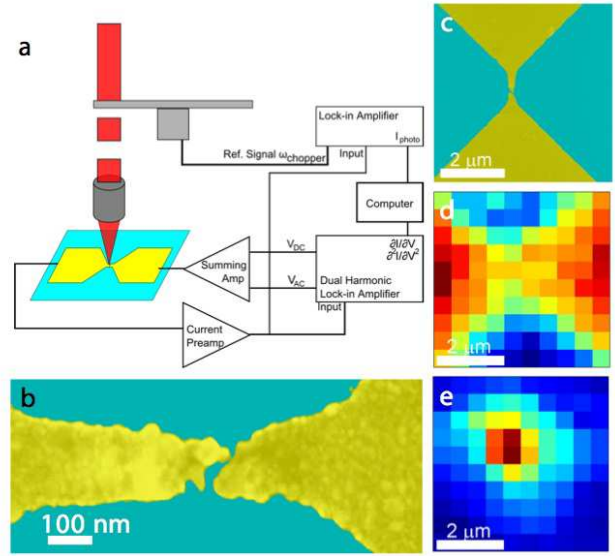}
\end{center}
\caption{\label{figSchematic}  Measurement approach and layout.
a) Schematic of electrical characterization measurement.  b) Colorized SEM image of typical nanogap device.  The actual tunnelling gap is not resolvable by SEM.  c) Larger SEM view of electrodes and nanogap. d, e)  False-colour maps of Si Raman line (d) and the photocurrent (e) for the device shown in (c).  In (d) blue indicates a weak Si signal, thus indicating the location of the Au electrodes. In (e) the color bar runs from red (20~nA) to blue (0~nA).  The photocurrent is clearly localized to the nanogap, to within optical resolution.    
}
\end{figure}

\clearpage

\begin{figure}
\begin{center}
\includegraphics[width=8.0cm,clip]{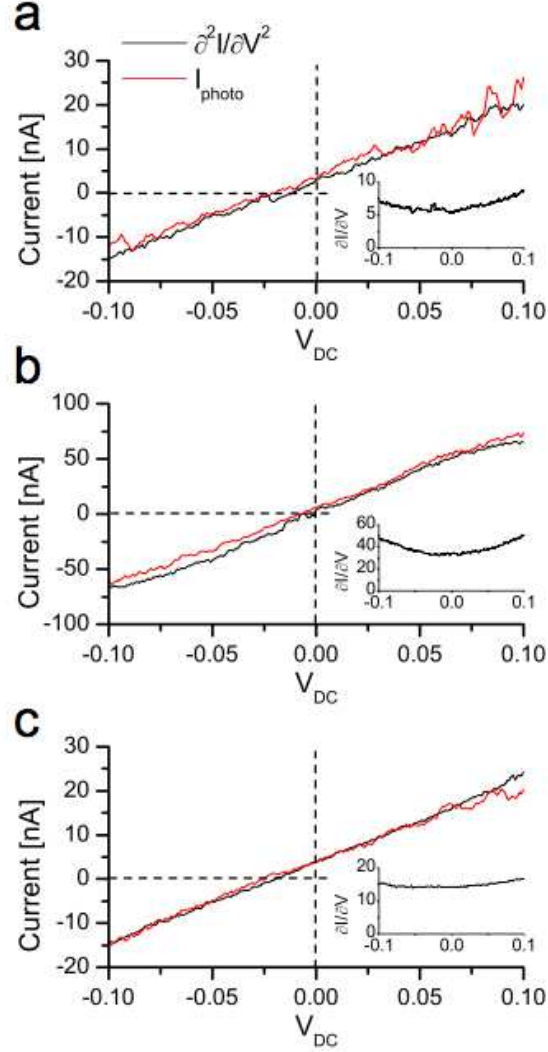}
\end{center}
\caption{\label{figIVCurves}  Demonstration of optical rectification.
a-c) Photocurrent (red, $I_{\mathrm{photo}}$) and $\frac{1}{4}V_{\mathrm{AC}}^{2}\partial^{2}I/\partial V^2$ (black) as a function of $V_{\mathrm{DC}}$ for three different samples.  The shared shapes of these curves, including changes of sign, demonstrate that the photocurrent arises from the rectification process.  Note the differing current scales for the three devices.  The different conductances and nonlinearities presumably result from microscopic details of the electrode surfaces and geometry.  Insets:   Conductance in units of $\mu$A/V for each device.  Gap distances from tunnelling measurements are 1.4~\AA, 0.44~\AA \ and 0.92~\AA.  Inferred $V_{\mathrm{opt}}$ values are 30~mV, 25~mV, and 33~mV, with uncertainties of 10\%.  Inferred electromagnetic fields are $2.1 \times 10^{8}$~V/m, $5.7 \times 10^8$~V/m and $3.6 \times 10^8$~V/m yield field enhancements of 718, 1940, and 1230, respectively.  
}
\end{figure}

\clearpage

\begin{figure}
\begin{center}
\includegraphics[width=8.5cm, clip]{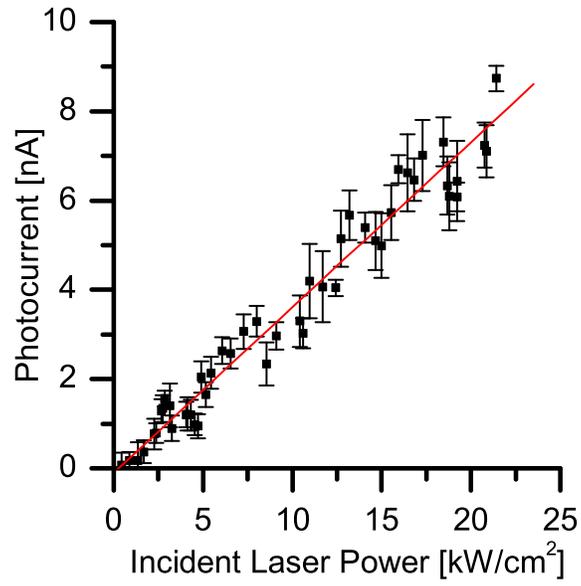}
\end{center}
\caption{\label{figPowerDependence}  Further evidence for optical rectification.
The rectified photocurrent, $I_{\mathrm{photo}}$, as a function of the incident laser power.  The error bars indicate one standard deviation of the $I_{\mathrm{photo}}$ measurements at each laser power.  The linear power dependence (see text) was found by a weighted least squares fit (red line).
}
\end{figure}

\begin{figure}
\begin{center}
\includegraphics[width=13cm, clip]{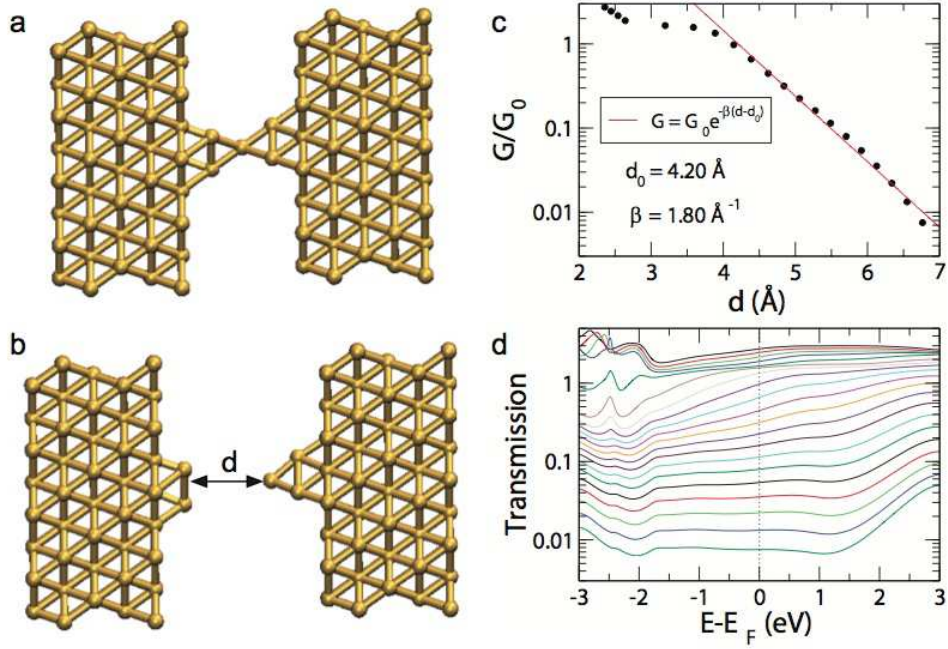}
\end{center}
\caption{\label{figTheory}  Theoretical basis for validity of rectification.
a) Geometry considered in the theoretical analysis of the distance dependence
of the linear conductance. We start with a single-atom contact grown along 
the $\langle 111 \rangle$ direction. b) Stretched geometry with a distance $d$
between the gold tips. c) Calculated linear conductance as a function of the
interelectrode distance (full circles). The solid line shows the fit to the exponential 
function $G=G_0 \exp[-\beta (d-d_0)]$. The fit parameters are indicated in the 
graph. d) Zero-bias transmission as a function of the energy for the different 
geometries of panel (c), with $d$ increasing from top to bottom.  Note the logarithmic vertical axis, and that the transmission remains a smooth function of energy over a broad range around the Fermi level.
}
\end{figure}

\begin{figure}
\begin{center}
\includegraphics[width=8.5cm, clip]{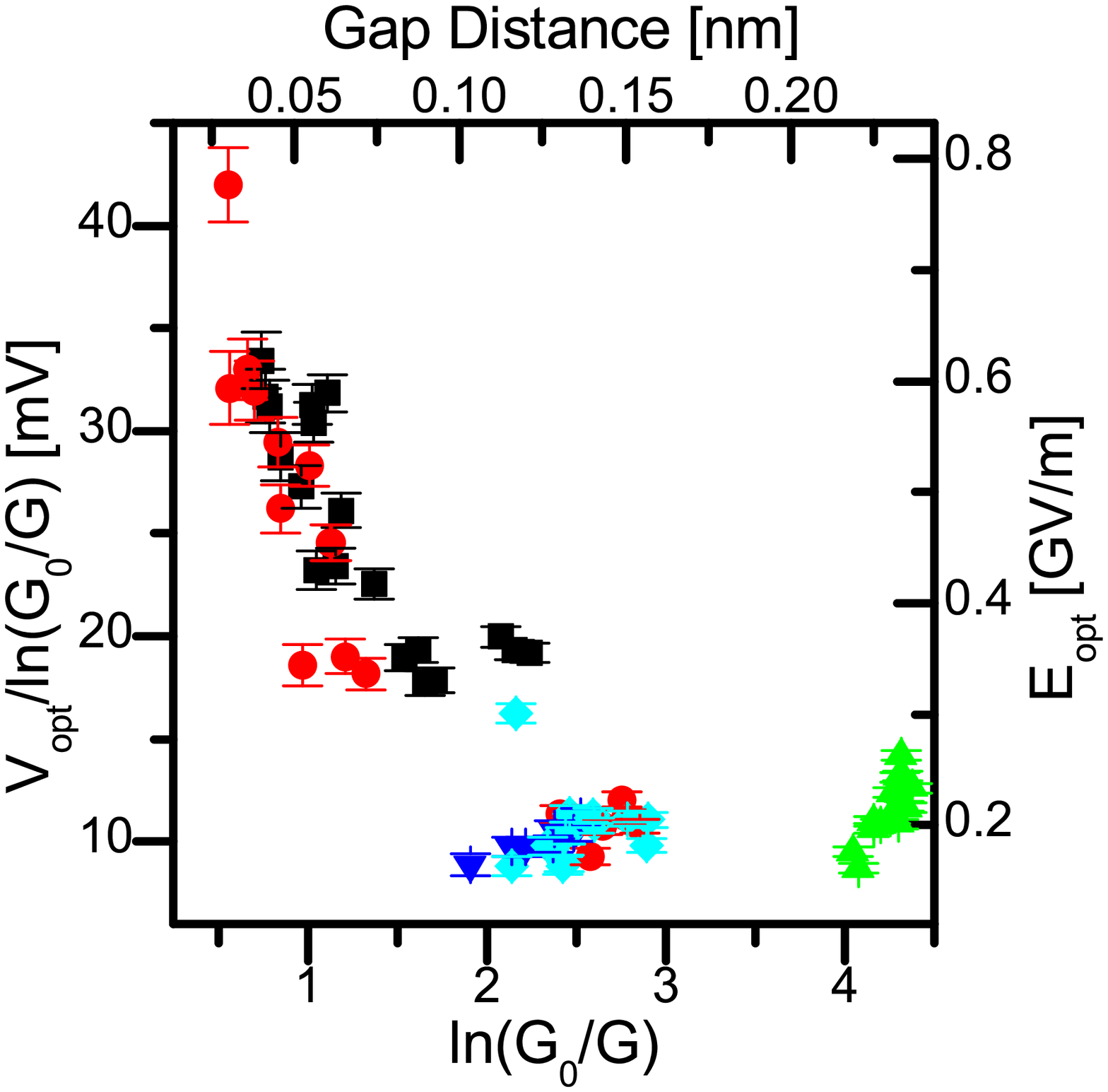}
\end{center}
\caption{\label{figdistdep}  Field enhancement (left axis) at the tunnelling region as a function of gap distance (top axis) for five devices (shown in different colours) measured a number of times at 80~K; the gap distance (and therefore the enhancement) changes as the gap geometry evolves over the course of repeated measurments.  The enhancement and gap distance are derived from measurements of the conductance (bottom axis shows $\ln(G_{0}/G)$) and the optical frequency voltage ($V_{\mathrm{opt}}$, right axis) based on the assumption that $d-d_{0}$ is the relevant scale over which $V_{\mathrm{opt}}$ falls, and $\beta = 1.85$~\AA$^{-1}$.  Error bars indicate the statistical uncertainty in $V_{\mathrm{opt}}$ and $\ln G$.
}
\end{figure}

\clearpage

{\center{\bf Supporting Information: \\ Optical Rectification and Field Enhancement in a Plasmonic Nanogap}}

\renewcommand{\thefigure}{S\arabic{figure}}
\setcounter{figure}{0}

\section{Tunnelling and field distributions in nanogaps}

The experimentally extracted parameter in the rectification measurement is the optical frequency potential difference, $V_{\mathrm{opt}}$, experienced by the tunnelling electrons.  To go from this potential difference to an estimate of the enhanced electric field within the tunnelling gap, it is necessary to estimate the distance over which this voltage is dropped.  In the main manuscript we use $d-d_{0}$, the interelectrode spacing inferred from the (low frequency) tunnelling conductance.  The question arises, is this reasonable given that $V_{\mathrm{opt}}$ is oscillating at optical frequencies?  There are two relevant issues:  the local character of the tunnelling process, and the penetration of electric field (and hence voltage drops) into the metal electrodes.

The rectification process is sensitive to the chemical potential difference experienced by precisely the tunnelling electrons.  Because of the local nature of the tunnelling process, the chemical potential difference we infer experimentally is measured between the "initial" and "final" positions of the tunnelling electrons.  The inferred $V_{\mathrm{opt}}$ is the voltage difference that drives the (extremely local) tunnelling process.  It is self-consistent to use the gap size extracted from tunnelling conductance as the relevant distance scale, since the local character of the tunnelling electrons is not expected to be frequency dependent (that is, there is no expectation that tunnelling takes place over longer distances at higher frequencies).   Additional voltage drops on longer scales within the metal electrodes are perfectly allowed, but not relevant to the inferred $V_{\mathrm{opt}}$ since the local tunnelling process is not sensitive to them.  The observed consistency between $I_{\mathrm{photo}}$ and $d^{2}I/dV^{2}$ over the whole bias range would be very difficult to understand if the local nature of the tunnelling process was fundamentally different between low and optical frequencies. 
This is certainly a subtle issue.  The definition of local electrostatic potential and its relationship with chemical potential in nanoscale junctions is not simple$^{\mathrm{S1}}$, since electronic distributions are driven out of equilibrium on a distance scale comparable to the inelastic scattering length.  A full treatment of optical-frequency tunnelling with realistic electronic structure of the contacts (well beyond the scope of the present work) would be needed to resolve this issue unambiguously.  The experimentally determined quantities in the main manuscript (tunnelling conductance, $V_{\mathrm{opt}}$) could be compared with more sophisticated calculations in the future.

Regarding field penetration into the metal, care is also required.   
The skin depth, which describes the characteristic attenuation length for the electric field of a plane wave normally incident on a metal surface, is related to the complex dielectric function at the incident frequency.  For gold at the incident wavelength of 785~nm, the skin depth is approximately 25~nm.
However, the skin depth calculation ignores collective electronic response such as plasmons.  In fact, in plasmonic nanostructures, while the field due to the incident radiation does penetrate the metal on length scales comparable to the skin depth, the total electric field \emph{including that produced by the plasmon response of the electron fluid} can be confined much more strongly than this.  

\begin{figure}[h]
  \begin{center}
    \includegraphics[width=8cm,clip]{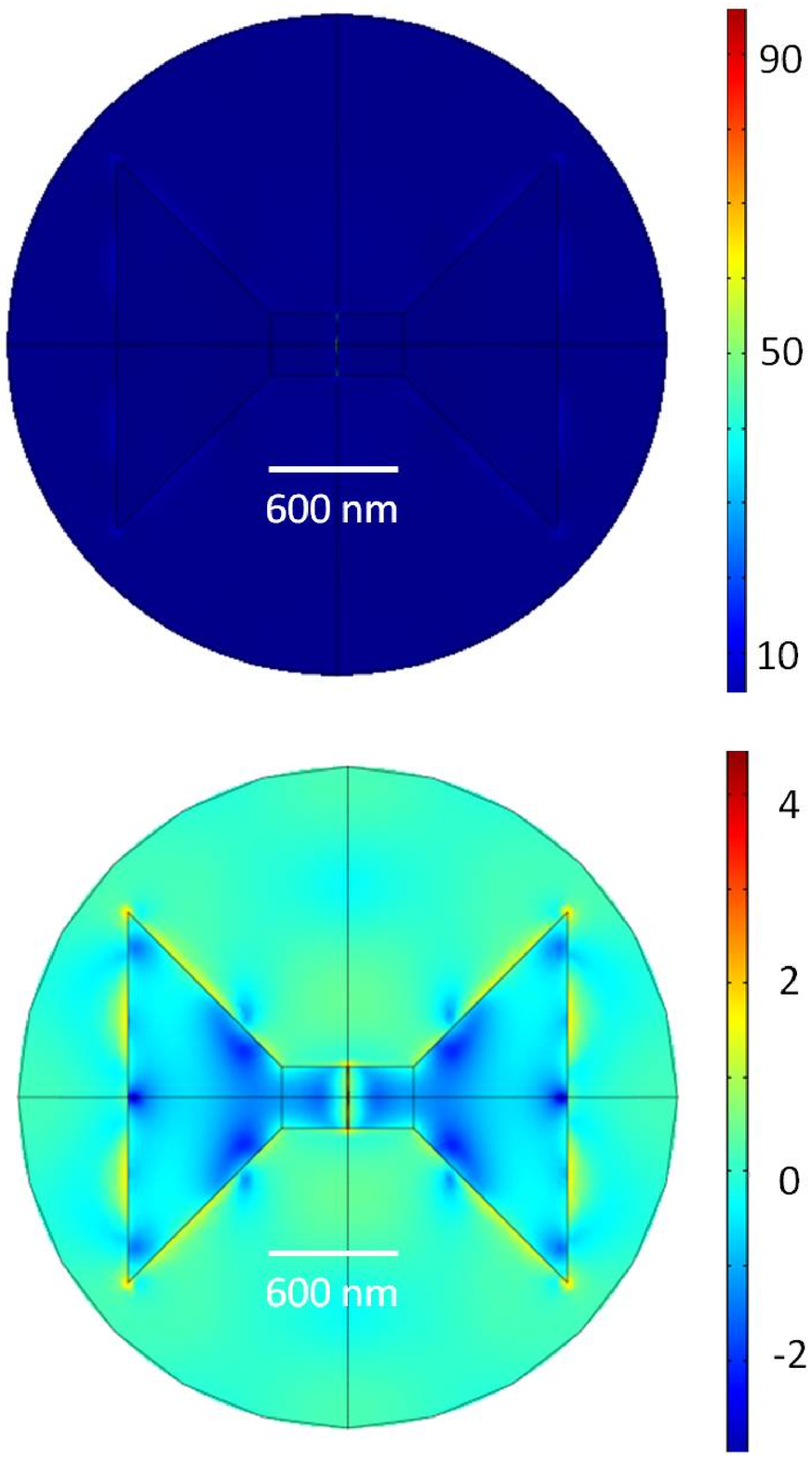}
  \end{center}
  \caption{\label{fig:fieldsfig1}Finite element simulation of the field distribution in a gold nanogap structure illuminated at its nanogap plasmon resonance.  Model parameters are described in the Supplemental Information text.  A spatial scale bar is shown.  The upper figure shows the magnitude of the local electric field normalized by the magnitude of the incident field on a \emph{linear} color scale.  Maximum field enhancement is by a factor of 100, at the closest interelectrode separation.  The lower figure shows the same data plotted with a (natural) logarithmic color scale. }

\end{figure}

\begin{figure}[h]
  \begin{center}
    \includegraphics[width=10cm,clip]{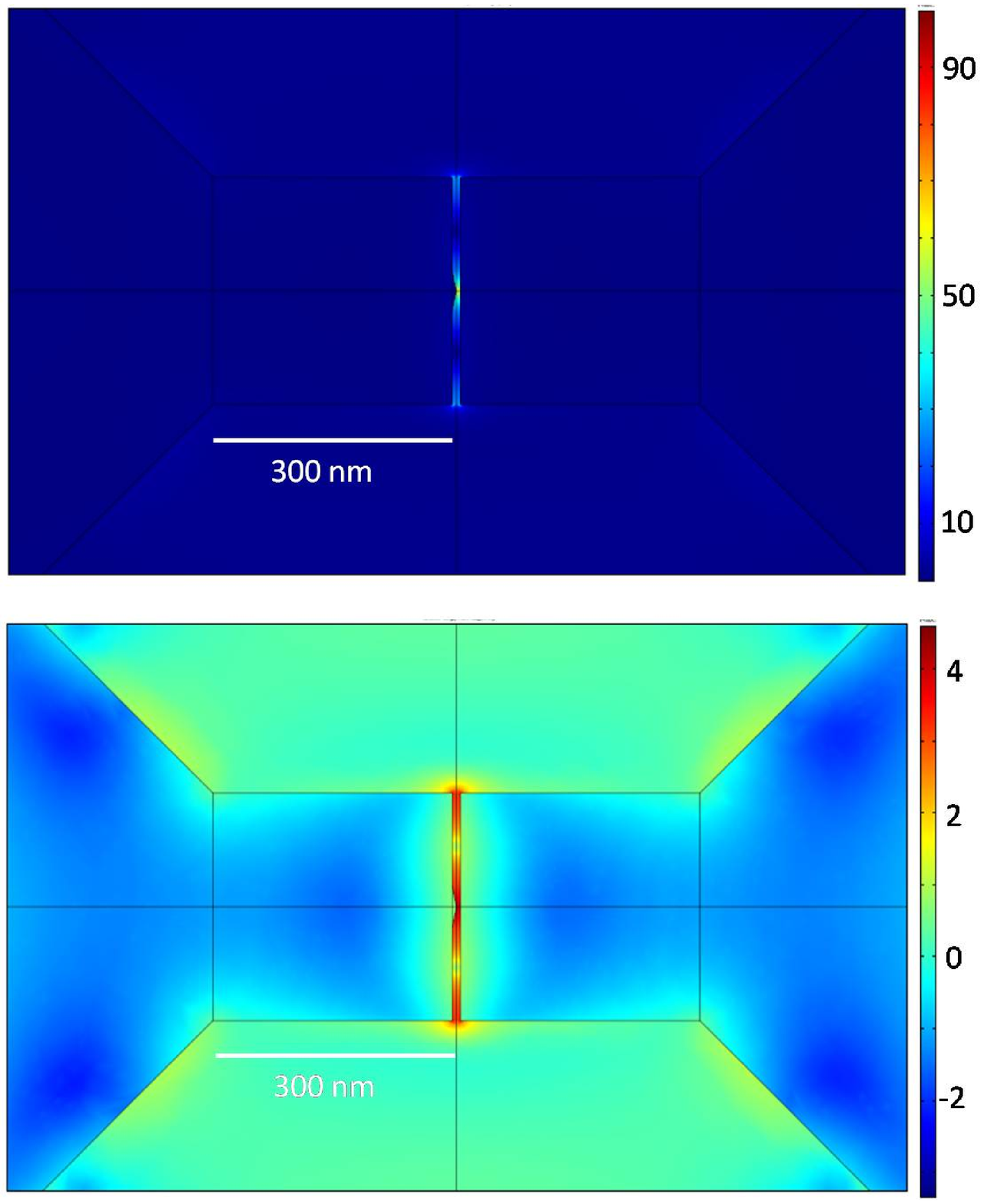}
  \end{center}
  \caption{\label{fig:fieldsfig2}A closer view of the model of Fig.~\ref{fig:fieldsfig1}.  While there is some electric field penetration into the metal, the vast majority of the total electric field is confined to the vacuum region between the electrodes.  That is, almost none of the potential difference between the electrodes is dropped within the metal.  Note further that the tunnelling process is sensitive only to the potential difference extremely local to the tip atoms.   }
\end{figure}

\begin{figure}[h]
  \begin{center}
    \includegraphics[width=10cm,clip]{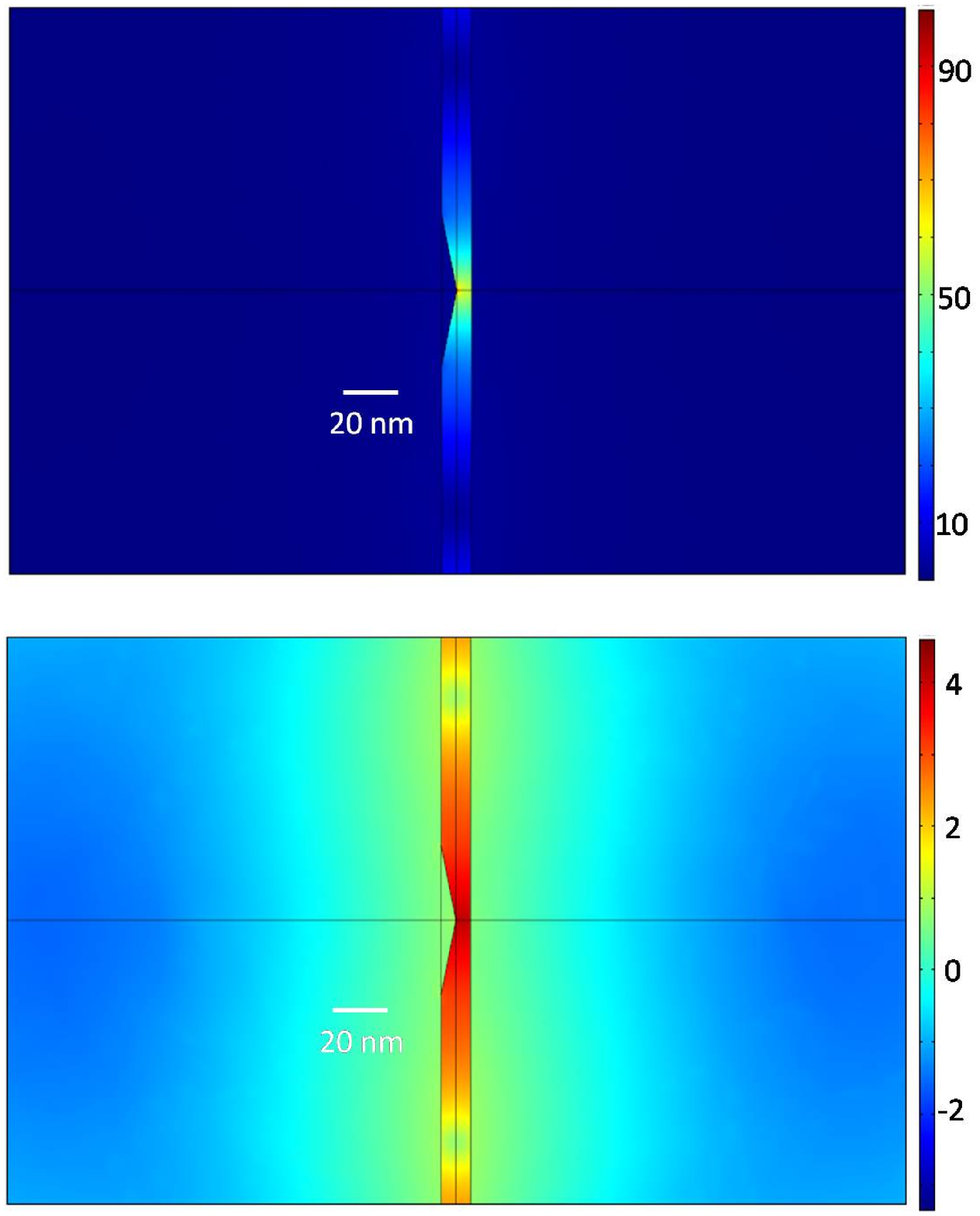}
  \end{center}
  \caption{\label{fig:fieldsfig3}A closer view of the model of Fig.~\ref{fig:fieldsfig2}.  Again, while there is some electric field penetration into the metal, the vast majority of the total electric field is confined to the vacuum region between the electrodes.  That is, almost none of the potential difference between the electrodes is dropped within the metal.  The skin depth parameter is \emph{not} relevant in determining this, because the skin depth does not account for fields produced by collective plasmon response of the electron liquid.   }
\end{figure}

Supplemental Figures \ref{fig:fieldsfig1}-\ref{fig:fieldsfig3} demonstrate this explicitly.  These figures show the results of a finite element simulation (using the commercial package COMSOL) of a model nanogap structure illuminated at the nanogap plasmon resonance.  The figures show top views of the metal structure at three different magnifications indicated by the scale bars.  The metal is modeled as Au (using the commonly employed Johnson and Christy dielectric function$^{\mathrm{S2}}$) 25~nm in thickness (chosen for computational efficiency).  The nanogap separation is 10~nm, except in the center of the device, where a protrusion on the left-hand electrode reduces the gap to 5~nm.  No dielectric substrate is included, but other models including such a substrate give qualitatively identical results.  The incident radiation is assumed to be a plane wave at normal incidence with a free space wavelength of 850~nm and a polarization directed along the line connecting the two electrodes (left-right in the figures).  Upper figures show the magnitude of the total electric field (normalized by the incident amplitude) on a linear color scale, while lower figures use a (natural) logarithmic color scale.  

It is immediately clear that the vast majority of the potential drop between the two electrodes takes place across the vacuum at the center of the nanogap.  Electric field does extend into the metal, but it is smaller than the nanogap field, by a factor of more than 50 in this case, a larger difference than one would expect based on a skin depth analysis that ignores plasmonic effects.  For smaller electrode separations, this disparity grows even further, with essentially 100\% of the (plasmon-dominated) optically induced potential difference dropping across the vacuum.  

The situation is, of course, complicated in the real devices by the true electronic structure of the metal surfaces at the atomic scale, and tunnelling conduction.    For very small gaps and distance scales, one must not trust the quantitative accuracy of finite-element or finite-difference time-domain calculations based on idealizations (infinitely sharp dielectric interfaces; bulk-like dielectric functions for the electrons and background charges at interfaces).  

However, in plasmon resonant structures such as these, the fact remains that a very large fraction of the potential drop takes place between the electrodes, and the skin depth is \emph{not} the relevant length scale to describe electric field confinement.  The excitation of plasmon collective modes strongly perturbs the charge density extremely local to the surface, leading to field  distributions that can differ greatly from those inferred in the absence of such collective effects.  For example, Liebsch$^{\mathrm{S3,S4}}$ shows that the presence of a surface plasmon excitation in a silver film or nanoparticle dynamically leads to surface charge density within a couple of atomic radii of the surface, with a length scale determined by quantum mechanical electronic structure effects and {\it not} by macroscopic electrodynamic effects such as the classical skin depth.

\section{Theoretical methods}

Here we describe in more detail the theoretical methods used for the
computation of the structural, electronic, and electrical properties of gold
nanocontacts. Moreover, we present an analysis of the geometry dependence of
the conductance decay with interelectrode distance.

\subsection{Electronic structure and contact geometries \label{subsec:elStrucGeo}}

For the determination of the electronic structure and geometry of the gold
contacts we employ density functional theory (DFT) as implemented in the
RI-DFT module of the quantum chemistry package {\it Turbomole} 5.9
$^{\mathrm{S5,S6}}$. In all our calculations we use the
open-shell formalism, the exchange-correlation functional BP86
$^{\mathrm{S7,S8}}$, and {\it Turbomole}'s standard Gaussian basis
set ``def-SVP'' of split-valence quality with polarization functions
$^{\mathrm{S6,S9,S10}}$.  Total energies are converged
to a precision of better than $10^{-6}$ atomic units, and structure
optimization is carried out until the maximum norm of the Cartesian gradients
has fallen below values of $10^{-4}$ atomic units.

An example of the contact geometries considered in this work is shown in
Fig.~\ref{fig:Au-contact}A, which corresponds to the junction of Fig.~4 in
the manuscript. In this case, a single-atom contact is
modeled by means of the extended central cluster (ECC) with 63 atoms
displayed in Fig.~\ref{fig:Au-contact}B. It consists of two pyramids grown
along the $\langle 111 \rangle$ direction of a fcc lattice, which contain,
beside a common tip atom, four layers with 3, 6, 10, and 12 atoms. We formally
divide the contact geometry into left ($L$), central ($C$), and right ($R$)
parts, and relax the atomic positions of the atoms in the $C$ region,
i.e.\ those of the tip atom and the neighboring layers with three atoms. In
contrast, the atoms in the $L$ and $R$ regions are kept fixed, and the lattice
constant is set to $4.08$ \AA, the experimental bulk value for gold.

In order to simulate what happens when the wires are broken and when one
enters the tunnelling regime, we stretch the ECC (Fig.~\ref{fig:Au-contact}B)
by separating the $L$ and $R$ regions in steps of 0.4 a.u. ($\approx 0.21$
\AA). After every stretching step, the geometries are reoptimized by relaxing
the $C$ region.

\subsection{Conductance calculations}

\begin{figure}[h]
  \begin{center}
    \includegraphics[width=14cm,clip]{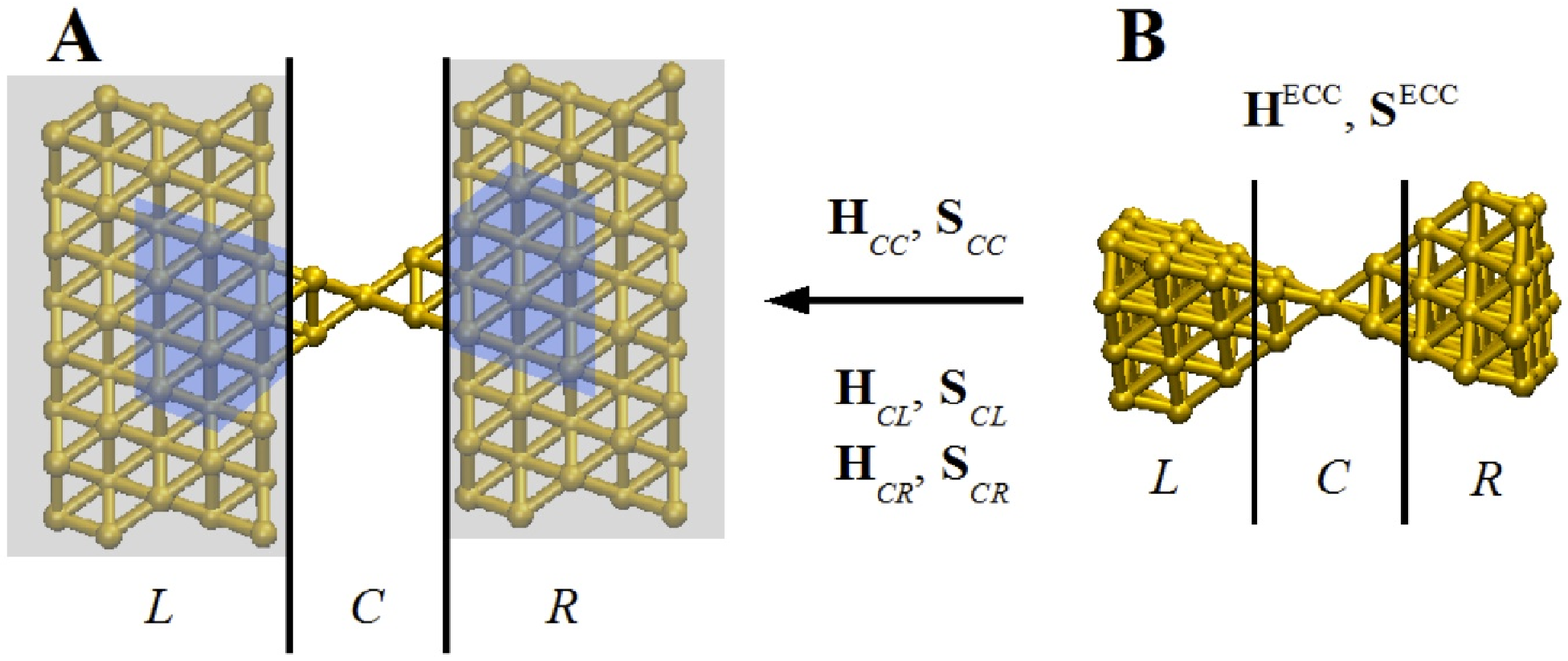}
  \end{center}
  \caption{\label{fig:Au-contact}Quantum transport scheme. The atomic-sized
    contact (A) is divided into a $C$ region and two semi-infinite $L$ and $R$
    electrodes. Using a similar division for the ECC (B), information on the
    electronic structure of the $C$ region (${\bf H}_{CC}, {\bf S}_{CC}$) as
    well as the $CL$ and $CR$ couplings (${\bf H}_{CL}, {\bf S}_{CL}$ and
    ${\bf H}_{CR}, {\bf S}_{CR}$) is extracted. The electrode surface Green's
    functions ${\bf g}^r_{LL} $ and ${\bf g}^r_{RR}$, needed for the
    computation of the self-energies ${\bf \Sigma}^r_L$ and ${\bf \Sigma}^r_R
    $ [Eq.~(\ref{eq:Sigma_X})], are determined in a separate calculation as 
    explained briefly in the text and in detail in Ref.~\cite{Pauly2008}.}
\end{figure}

To compute the charge transport properties we apply a method based on standard
Green's function techniques and the Landauer formula expressed in a local
nonorthogonal basis, as described in Ref.~\cite{Pauly2008}. Briefly, the local
basis allows us to partition the basis states into $L$, $C$, and $R$ ones,
according to the division of the contact geometry
(Fig.~\ref{fig:Au-contact}). Thus, the Hamiltonian (or single-particle Fock)
matrix ${\bf H}$, and analogously the overlap matrix ${\bf S}$, can be written
in the block form
\begin{equation}
  {\bf H}=\left(\begin{array}{ccc}
    {\bf H}_{LL} & {\bf H}_{LC} & {\bf 0}\\
    {\bf H}_{CL} & {\bf H}_{CC} & {\bf H}_{CR}\\
    {\bf 0} & {\bf H}_{RC} & {\bf H}_{RR}\end{array}\right).\label{eq:H}
\end{equation}
Within the Landauer approach, the low-temperature conductance is given by $G =
G_0 \tau(E_{\rm F})$, where $G_0=2e^2/h$ and $\tau(E_{\rm F})$ is the
transmission of the contact evaluated at the Fermi energy.  The
energy-dependent transmission $\tau(E)$ can be expressed in terms of the
Green's functions as$^{\mathrm{S1}}$:
\begin{equation}
  \tau(E) = \mathrm{Tr} \left[{\bf \Gamma}_{L} {\bf G}_{CC}^{r} {\bf
      \Gamma}_{R} {\bf G}_{CC}^{a} \right],\label{eq:TE}
\end{equation}
where the retarded Green's function is given by
\begin{equation}
  {\bf G}_{CC}^{r}(E) = \left[ E {\bf S}_{CC} - {\bf H}_{CC} -
    {\bf \Sigma}_{L}^{r}(E) - {\bf \Sigma}_{R}^{r}(E)\right]^{-1} ,
  \label{eq:G_CC}
\end{equation}
and ${\bf G}_{CC}^{a} = \left[{\bf G}_{CC}^{r}\right]^{\dagger}$.  The
self-energies adopt the form
\begin{equation}
  {\bf \Sigma}_{X}^{r}(E) = \left({\bf H}_{CX} - E {\bf S}_{CX} \right) {\bf
    g}_{XX}^{r}(E) \left({\bf H}_{XC} - E {\bf S}_{XC} \right),
  \label{eq:Sigma_X}
\end{equation}
the scattering rate matrices are given by ${\bf \Gamma}_{X}(E) = -2\mathrm{Im}
\left[{\bf \Sigma}_{X}^{r}(E) \right]$, and ${\bf g}_{XX}^{r}(E)=(E {\bf S}_{XX}
- {\bf H}_{XX})^{-1}$ are the electrode Green's functions with $X=L, R$.

In order to describe the transport through the contact shown in
Fig.~\ref{fig:Au-contact}A, we first extract ${\bf H}_{CC}$ and
${\bf S}_{CC}$ and the matrices ${\bf H}_{CX}$
and ${\bf S}_{CX}$ from a DFT calculation of the ECC in
Fig.~\ref{fig:Au-contact}B. The blue-shaded atoms in regions $L$ and $R$ of
Fig.~\ref{fig:Au-contact}A are assumed to be those coupled to the $C$ region.
Thus, ${\bf H}_{CX}$ and ${\bf S}_{CX}$, obtained from the ECC, serve as the
couplings to the electrodes in the construction of ${\bf \Sigma}_{X}^{r}(E)$.

On the other hand, the electrode Green's functions ${\bf g}_{XX}^{r}(E)$ in
Eq.~(\ref{eq:Sigma_X}) are modeled as surface Green's functions of ideal
semi-infinite crystals. In order to obtain them, we have computed separately
the electronic structure of a spherical fcc gold cluster with 429 atoms. By
extracting the Hamiltonian and overlap matrix elements connecting the atom in
the origin of the cluster with all its neighbors and by using these as ``bulk
parameters'', we construct a semi-infinite crystal which is infinitely
extended perpendicular to the transport direction. The surface Green's
functions are then calculated from this crystal with the help of the
decimation technique $^{\mathrm{S11,S12}}$. In this way we describe the
whole system consistently within DFT, using the same nonorthogonal basis set
and exchange-correlation functional everywhere.

We assume the Fermi energy $E_{\rm F}$ to be fixed by the gold leads.  From the
Au$_{429}$ cluster we obtain $E_{\rm F}=-5.0$ eV. Let us point out that the Fermi
energy is used only to read off the conductance from the transmission function
$\tau(E)$, but it does not enter into our calculations otherwise.

\subsection{Geometry dependence of the conductance decay with 
interelectrode distance}

When an atomic contact is broken, the conductance depends exponentially on the
interelectrode distance $d$, i.e.\ $G=G_0 \exp(-\beta (d-d_0))$. Here, $d_0$
describes the prefactor of the exponential law and $\beta$ is the attenuation
factor. Both constants depend on the surface morphology, in particular on the
local atomic environment at the interface region as well as on the
crystallographic direction along which the contact is grown. We have studied
the influence of these two factors by performing different DFT-based simulations 
of the breaking process of gold nanowires.

\begin{figure}[h]
  \begin{center}
    \includegraphics[width=12cm,clip]{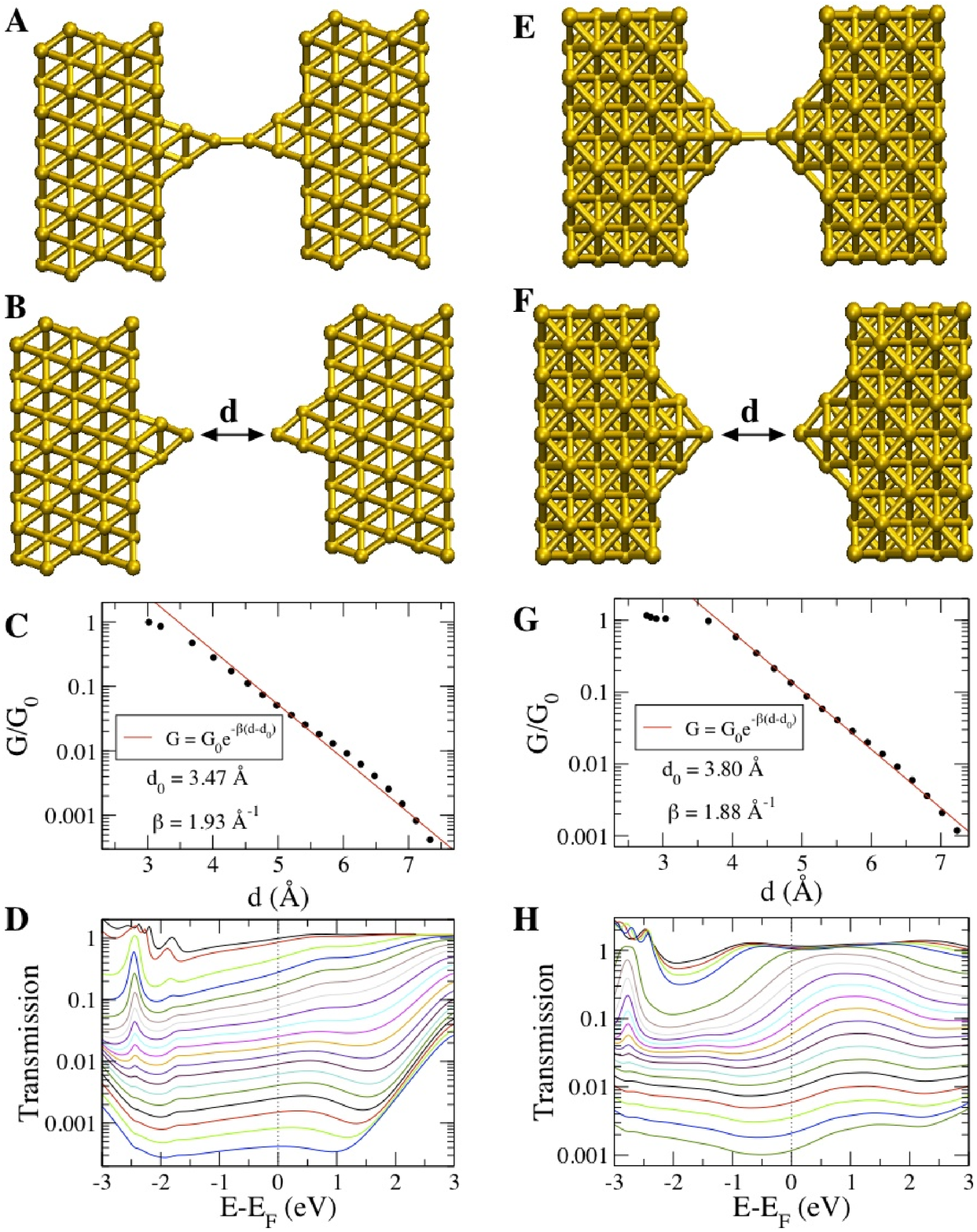}
  \end{center}
  \caption{\label{fig:beta-factor} (A) Starting geometry used to study the
    distance dependence of the conductance of a gold contact. This geometry
    corresponds to a one-atom-thick contact grown along the $\langle 111
    \rangle$ direction of a fcc lattice with a dimer structure (or chain of
    two atoms) in the middle. (B) Stretched geometry with the distance $d$
    between the gold tips. (C) Calculated linear conductance as a function of
    the interelectrode distance (full circles). The solid line shows the fit
    to the exponential function $G=G_{0} \exp[-\beta (d-d_{0})]$. The fit
    parameters $d_0$ and $\beta$ are indicated in the graph. (D) Zero-bias
    transmission $\tau(E)$ as a function of the energy (measured with respect
    to the Fermi energy) for the geometries of panel C. (E-H) The same as in
    panels A-D for a dimer contact grown along the $\langle 100 \rangle$
    direction.}
\end{figure}

In Fig.~\ref{fig:beta-factor}A-D we show the results obtained for a contact
grown along the $\langle 111 \rangle$ direction. As compared to the geometry
considered in Fig.~4 of the manuscript, the starting geometry is a
one-atom-thick contact with a dimer (or two-atom chain) in the narrowest
part. We have stretched this contact following the procedure described above
(see paragraph ``Electronic structure and contact geometries''), and we have
computed both the linear conductance and the energy-dependent transmission at
the different stages of the elongation process. As it is shown in
Fig.~\ref{fig:beta-factor}C, in the range relevant for the experiments, the
conductance decays exponentially with an exponent $\beta =
1.93$~\AA$^{-1}$. This value is approximately 7\% larger than the value
obtained for the contact of Fig.~4. Furthermore, in
Fig.~\ref{fig:beta-factor}D one can see that the transmission is still a
smooth function of the energy in a range larger than 1 eV around the Fermi
energy.

In Fig.~\ref{fig:beta-factor}E-H we present the corresponding results for a
similar contact with a dimer structure, but this time grown along the $\langle
100 \rangle$ direction of the fcc lattice. In this case $\beta$ takes a value
of 1.88~\AA$^{-1}$, which is 4\% larger than those found in Fig.~4. Again,
the transmission is a rather featureless function of the energy in the relevant
conductance range.

To conclude, we have analyzed different types of gold atomic-sized
contacts. Our results indicate that the attenuation factor $\beta$ varies
roughly within 10\% of the value reported in the manuscript. Additionally,
they corroborate that the transmission is a rather smooth function of the
energy in an interval of approximately 2 eV around the Fermi energy for the
tunnelling regime explored in the experiments.


\clearpage

\clearpage

\clearpage

\clearpage

\clearpage

\clearpage

\noindent [S1]  Datta, S. {\it Electronic Transport in Mesoscopic Systems}. Cambridge Studies in Semiconductor
Physics and Microelectronic Engineering (Cambridge University, 1995).

\noindent [S2] Johnson, P. B. \& Christy, R. W. Optical Constants of the Noble Metals. {\it Phys. Rev. B} {\bf 6},
4370-4379 (1972).

\noindent [S3] Liebsch, A. Surface plasmon dispersion of Ag. {\it Phys. Rev. Lett.} {\bf 71}, 145-148 (1993).

\noindent [S4] Liebsch, A. Surface-plasmon dispersion and size dependence of Mie resonance: Silver
versus simple metals. {\it Phys. Rev. B} {\bf 48}, 11317-11328 (1993).

\noindent [S5] Ahlrichs, R., B{\"a}r, M., H{\"a}ser, M., Horn, H. \& K{\"o}lmel, C. Electronic structure calculations
on workstation computers: The program system turbomole. {\it Chem. Phys. Lett.} {\bf 162}, 165-169 (1989).

\noindent [S6] Eichkorn, K., Treutler, O., {\"O}hm, H., H{\"a}ser, M. \& Ahlrichs, R. Auxiliary basis sets to
approximate Coulomb potentials. {\it Chem. Phys. Lett.} {\bf 242}, 652-660 (1995).

\noindent [S7] Becke, A. D. Density-functional exchange-energy approximation with correct asymptotic
behavior. {\it Phys. Rev. A} {\bf  38}, 3098-3100 (1988).

\noindent [S8]  Perdew, J. P. Density-functional approximation for the correlation energy of the inhomogeneous
electron gas. {\it Phys. Rev. B} {\bf 33}, 8822-8824 (1986).

\noindent [S9]  Sch{\"a}fer, A., Horn, H. \& Ahlrichs, R. Fully optimized contracted Gaussian basis sets for
atoms Li to Kr. {\it J. Chem. Phys.} {\bf  97}, 2571-2577 (1992).

\noindent [S10] Eichkorn, K., Weigend, F., Treutler, O. \& Ahlrichs, R. Auxiliary basis sets for main
row atoms and transition metals and their use to approximate Coulomb potentials. {\it Theor.
Chem. Acc.} {\bf  97}, 119-124 (1997).

\noindent [S11]  Pauly, F. {\it et al.} Cluster-based density-functional approach to quantum transport through
molecular and atomic contacts. {\it New J. Phys.} {\bf 10}, 125019 (2008).

\noindent [S12] Guinea, F., Tejedor, C., Flores, F. \& Louis, E. Effective two-dimensional Hamiltonian at
surfaces. \it{Phys. Rev. B} {\bf 28}, 4397-4402 (1983).

\end{document}